\documentclass[twocolumn,floatfix, showpacs]{revtex4}

\usepackage[final]{epsfig}
\usepackage{amsmath}
\usepackage{parskip}

\begin{document}

\title{Flow dependence of high $p_T$ parton energy loss in heavy-ion collisions}

\author{Thorsten Renk, J\"{o}rg  Ruppert}


\affiliation{Department of Physics, Duke University, PO Box 90305,  Durham, NC 27708 , USA}

\begin{abstract}
The measured transverse momentum spectra and HBT correlations of bulk (i.e. low $p_T$) matter can be well explained by assuming that the soft sector of particles produced in ultrarelativistic heavy-ion collisions is (approximately) thermalized and undergoes collective accelerated expansion in both longitudinal and transverse direction. However, this implies that bulk matter will have a non-vanishing flow component transverse to the trajectory of a high $p_T$ partonic jets. In general, this will increase the energy loss experienced by the jet parton and modify the shape of the jet cone. In this paper, we present a systematic study of the magnitude of the additional energy loss induced by flow under realistic assumptions for the medium evolution. We argue that a perturbative QGP description may be sufficient for the measured $R_{AA}$ if flow during the medium evolution is taken into account properly.

\end{abstract}

\maketitle

\section{Introduction}
\label{sec_introduction}

Energy loss of a high $p_T$ 'hard' parton travelling through low $p_T$ 'soft' matter has long been recognized as a promising tool to study the initial high-density phases of ultrarelativistic
heavy-ion collisions (URHIC) \cite{Jet0, Jet00,Jet1,Jet2,Jet3,Jet4,Jet5,Jet6}. In \cite{Urs1,Urs2}, it
has been suggested that a flow component transverse to the high $p_T$ parton trajectory would lead to increased energy loss as compared to the one in a static medium.

Descriptions of measured low $p_T$ spectra and HBT correlation radii for Au-Au collisions at RHIC energies in terms of a thermalized medium characterized by a collective expansion (flow) velocity field at thermal decoupling are quite successful, see e.g. \cite{LisaBlastWave}. 
Experimental measurements of charge independent two particle correlations seem to indicate that the distribution of jet secondaries is widened in longitudinal direction \cite{Minijets}. A possible explanation could be that longitudinal flow leaves a characteristic imprint on the energy loss phenomenon.

However, since jets propagate through the medium from the early onset of thermalization and lose energy mainly in the initial hot and dense phase, the knowledge of the flow profile at thermal decoupling is only of limited value. 
Instead, one has to study a more generalized framework which allows
to characterize the whole evolution of thermalized matter.
It is the aim of this paper to provide such a study in a framework which has been shown to lead to a good description of the experimental data of transverse mass spectra and HBT correlations \cite{RenkSpectraHBT} and which is flexible enough to test also various other evolution scenarios (which are at present not favoured by the data). In the present paper we focus on the energy loss of the leading jet particle and
leave predictions for the distortion of the jet cone to a subsequent publication.

\section{The formalism}

The parton's energy loss depends on the position of its production point $\vec{r}_0$ and the angular orientation of it's trajectory $\vec{r}$. 
In order to determine the probability for a hard parton $P(\Delta E)$ to lose the energy
$\Delta E$ while traversing the medium on it's trajectory, we make use of a scaling law \cite{JetScaling} which allows to relate the dynamical scenario to a static equivalent one whose density has been rescaled. This probability distribution of the energy loss $\Delta E$ is given as a functional of the distribution $\omega \frac{dI}{d\omega}$ of gluons emitted into the jet cone \cite{QuenchingWeights}:
\begin{widetext}
\begin{equation}
P(\Delta E) = \sum_{n=0}^\infty \frac{1}{n!} \left[ \prod_{i=1}^n \int d \omega_i \frac{dI(\omega_i)}{d \omega}\right]
\delta\left( \Delta E - \sum_{i=1}^n \omega_i\right) \exp\left[-\int d\omega\frac{dI}{d\omega} \right].
\end{equation}
\end{widetext}

The explicit expressions for this quantity are different if one calculates $\omega \frac{dI}{d\omega}$ under the assumption that multiple soft scattering processes or a few hard scattering processes are responsible for the energy loss. For both cases they can be found in \cite{QuenchingWeights}.
Using these results we obtain $P(\Delta E)$ from the key quantity of jet energy loss, namely the local transport coefficient $\hat{q}(\eta_s, r, \tau)$. 
The transport coefficient characterizes the squared average momentum transfer from the medium to the hard parton per unit pathlength. Since we consider a time-dependent inhomogeneous medium, this quantity depends on spacetime rapidity 
 rapidity $\eta_s = \frac{1}{2}\ln(t+z)/(t-z)$, radius $r$ and proper time $\tau = \sqrt{t^2-z^2}$. For simplicity we focus on central collisions and assume azimuthal symmetry. 
The transport coefficient is related to the energy density of the medium as
$\hat{q} = c \epsilon^{3/4}$. In the case of an ideal quark-gluon plasma (QGP) one would expect $c\approx2$ \cite{Baier}. 
It depends also on the jet trajectory $\vec{r}=\vec{r}_0+\xi \vec{n}$ (production point $\vec{r_0}$ and orientation $\vec{n}$).
In order to calculate the transport coefficient in the presence of flow, we follow the prescription suggested in \cite{Urs2} and replace
\begin{equation}
\hat{q} = c \epsilon^{3/4}(p) \rightarrow c \epsilon^{3/4} (T^{n_\perp  n_\perp})
\end{equation}
with
\begin{equation}
T^{n_\perp n_\perp} = p(\epsilon) + \left[ \epsilon + p(\epsilon)\right] \frac{\beta_\perp^2}{1-\beta_\perp^2},
\end{equation}
where $\beta_\perp$ is the spatial component of the flow field orthogonal to the parton trajectory. $T^{n_\perp n_\perp}$ indicates the component of the energy-momentum tensor, where $n_\perp$ is orthogonal to the jet's trajectory. In the absence of any flow effects the original result $\hat{q}=c\epsilon^{3/4}(p)$ is recovered. 

The transport coefficient enters the calculation of $P(\Delta E)$ via $\omega_c$ and $\langle \hat{q}L\rangle$, which are the linearly line-averaged characteristic gluon energy:
\begin{equation}
\omega_c({\bf r_0}, \phi) = \int_0^\infty d \xi \xi \hat{q}(\xi)
\end{equation}
 and the time-averaged total transverse momentum squared:
\begin{equation}
\langle\hat{q}L\rangle ({\bf r_0}, \phi) = \int_0^\infty d \xi \hat{q}(\xi),
\end{equation}
respectively  \cite{QuenchingWeights}.

The calculation of the transport coefficient and subsequent 
$\omega_c({\bf r_0}, \phi)$, $(\hat{q}L) ({\bf r_0}, \phi) $ and finally $P(\Delta E)$ requires a model of the fireball evolution which determines the local energy density and the flow profile. We discuss details of that model in the next section.
In our calculation we average over all possible angles and
production vertices, weighting the distribution of the jets with the nuclear overlap factor $T_{AA}({\bf b}) = \int dz \rho^2({\bf b},z)$
(for central collisions) with $\rho$ the nuclear density as a function of impact parameter {\bf b} and longitudinal coordinate $z$.

This formalism sets the stage for the calculation of the nuclear modification factor. In the case of central collisions it is given by
\begin{equation}
R_{AA}(p_T,y) = \frac{d^2N^{AA}/dp_Tdy}{T_{AA}(0) d^2 \sigma^{NN}/dp_Tdy}.
\end{equation}

To obtain $R_{AA}$ we use the formalism and averaging procedure as described above and calculate the inclusive charged hadrons production in LO pQCD.  This amounts to folding the average energy loss probability into the factorized expression for hadron production (for brevity we give schematical expressions, an explicit representation can be found in \cite{Kari1, Kari2}):

\begin{equation}
d\sigma_{med}^{AA\rightarrow h+X} = \sum_f d\sigma_{vac}^{AA \rightarrow f +X} \otimes P_f(\Delta E) \otimes
D_{f \rightarrow h}^{vac}(z, \mu_F^2)
\end{equation} 

where

\begin{equation}
d\sigma_{vac}^{AA \rightarrow f +X} = \sum_{ijk} f_{i/A}(x_1,Q^2) \otimes f_{j/A}(x_2, Q^2) \otimes \hat{\sigma}_{ij 
\rightarrow f+k}.
\end{equation}

Here, $f_{i/A}(x, Q^2)$ denotes the distribution of the parton $i$ inside the nucleus as a function of the parton
momentum fraction $x$ and the hard scale $Q^2$ of the scattering. Likewise, $D_{f \rightarrow \pi}^{vac}(z, \mu_F^2)$
denotes the fragmentation function of a parton $f$ into a hadron with the hadron taking the fraction $z$ of the parton momentum
and a fragmentation scale $\mu_f^2$ which should be a typical hadronic scale $O(1 \text{ GeV})$.
The expressions for the hard pQCD cross sections $\hat{\sigma}_{ij\rightarrow f+k}$ can e.g. be found in \cite{pQCD-Xsec}.
We use the CTEQ6 parton distribution functions \cite{CTEQ1, CTEQ2} for the pp reference, the
NPDF set \cite{NPDF} for production in nuclear collisions and the KKP fragmentation functions \cite{KKP}.

Unless stated otherwise, we assume for the strong coupling $\alpha_s = 0.3$ in the calculation of $P(\Delta E)$ in following.

\section{The fireball evolution model}

The fireball evolution enters this framework in the shape of $\epsilon(\eta_s,r,\tau)$ and the flow
profile $u^\mu(\eta_s,r,\tau)$. For the description of the evolution, we base our investigation on
the formalism outlined in \cite{RenkSpectraHBT}. 

The main assumption for the model is that an equilibrated system is formed
a short time $\tau_0$ after the onset of the collision. Furthermore, we assume that this
thermal fireball subsequently expands isentropically until the mean free path of particles exceeds
(at a time $\tau_f$) the dimensions of the system and particles 
move without significant interaction to the detector.

For the entropy density at a
given proper time we make the ansatz 
\begin{equation}
s(\tau, \eta_s, r) = N R(r,\tau) \cdot H(\eta_s, \tau)
\end{equation}
with $\tau $ the proper time as measured in a frame co-moving
with a given volume element  and $R(r, \tau), H(\eta_s, \tau)$ two functions describing the shape of the distribution
and $N$ a normalization factor.
We use Woods-Saxon distributions
\begin{equation}
\begin{split}
&R(r, \tau) = 1/\left(1 + \exp\left[\frac{r - R_c(\tau)}{d_{\text{ws}}}\right]\right)
\\ & 
H(\eta_s, \tau) = 1/\left(1 + \exp\left[\frac{\eta_s - H_c(\tau)}{\eta_{\text{ws}}}\right]\right).
\end{split}
\end{equation}
to describe the shapes for a given $\tau$. Thus, the ingredients of the model are the 
skin thickness parameters $d_{\text{ws}}$ and $\eta_{\text{ws}}$
and the para\-me\-tri\-zations of the expansion of the spatial extensions $R_c(\tau), H_c(\tau)$ 
as a function of proper time.  For a radially non-relativistic 
expansion and constant acceleration we find
$R_c(\tau) = R_0 + \frac{a_\perp}{2} \tau^2$. $H_c(\tau)$ is obtained
by integrating forward in $\tau$ a trajectory originating from the collision center which is characterized
by a rapidity  $\eta_c(\tau) = \eta_0 + a_\eta \tau$
with $\eta_c = \text{atanh } v_z^c$ where $v_z^c$ is  the longitudinal velocity of that 
trajectory. Since the relation between proper time as measured in the co-moving frame 
and lab time is determined by the rapidity at a given time, the resulting integral is
 solved numerically (see \cite{Synopsis} for details).
$R_0$
is determined in overlap calculations using Glauber theory, the initial size of the rapidity interval occupied
by the fireball matter. $\eta_0$ is a free parameter and we choose to use the transverse velocity  $v_\perp^f = a_\perp \tau_f$ and
rapidity at decoupling proper time $\eta^f = \eta_0 + a_\eta \tau_f$ as parameters.
Thus, specifying $\eta_0, \eta_f, v_\perp^f$ and $\tau_f$ sets the essential scales of the spacetime
evolution and $d_{\text{ws}}$ and $\eta_{\text{ws}}$ specify the detailed distribution of entropy density.

For transverse flow we assume a linear relation between radius $r$ and
transverse rapidity $\rho = \text{atanh } v_\perp(\tau) =  r/R_c(\tau) \cdot \rho_c(\tau)$
with $\rho_c(\tau) = \text{atanh } a_\perp \tau$. In the following investigation we also test a flow profile $\rho \sim r^2$.

We allow for the possibility of accelerated longitudinal expansion
which in general implies $\eta \neq \eta_s$ \cite{Synopsis}.  Here, $\eta = \frac{1}{2} \ln\frac{p_0 + p_z}{p_0 - p_z}$
denotes the longitudinal momentum rapidity of a given volume element. We can parametrize this mismatch between
spacetime and momentum rapidity as
a local $\Delta \eta = \eta -\eta_s$ which is a function of $\tau$ and $\eta_s$.

\section{Results}

In the following we calculate  $\omega \frac{dI}{d\omega}$ in the multiple soft scattering
approximation. The largest remaining uncertainty is the detailed choice of the parameter $c$ in the relation 
$\hat{q} = c \epsilon (T^{n_\perp  n_\perp})$ connecting transport coefficient and local energy density.
As the highest temperature reached in the model evolution is only approximately two times the phase transition
temperature $T_C$ there is no a priori reason to assume that the ideal QGP value $c\approx 2$ would be realized.
Hence, unless stated otherwise, we adjust $c$ such that the scenario {\it without flow} reproduces the
large $p_T$ region (we do not include any intermediate $p_T$ physics such as the Cronin enhancement in
this study) and discuss changes induced by flow relative to this baseline. 

\begin{figure}[!htb]
\epsfig{file=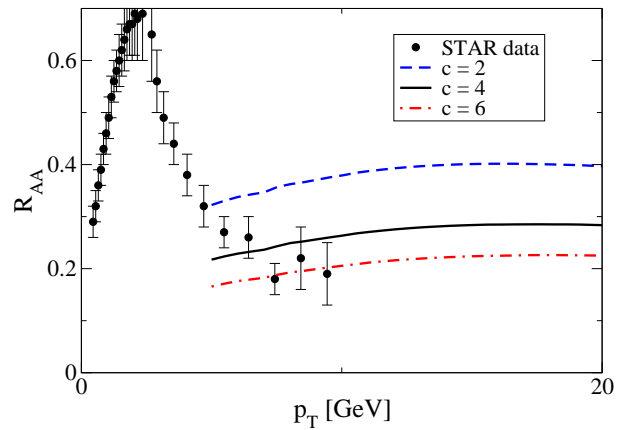, width=8cm}
\caption{\label{F-1}Calculated $R_{AA}$ as a function of transverse momentum for different values of the
proportionality constant $c$ between transport coefficient $\hat{q}$ and energy density $\epsilon^{3/4}$
compared with STAR data \cite{STAR_RAA}, all without taking the effect of flow to the energy loss into account. }
\end{figure}

As in \cite{Kari1} we find that $R_{AA}$ stays rather flat as a function of $p_T$ out to more than
50 GeV. If the flattening which may be seen in the data can be trusted, a value of $c=4-6$ would
lead to agreement with the data by the STAR collaboration for 5\% central 200 AGeV Au-Au collisions \cite{STAR_RAA}
(see Fig.~\ref{F-1}).
Since taking into account the effect of flow transverse to the jet axis is expected to increase energy loss, 
we choose the value $c=4$ as a baseline for the following investigation.

Taking the transverse flow field into account we indeed observe increased supression, i.e. a reduction of $R_{AA}$. Its magnitude 
shows some dependence on how the flow field depends on the radial position, we investigate
$\rho \sim r$ and $\rho \sim r^2$. In both cases the maximum transverse velocity is adjusted such that the measured
$\pi^-, K^-$ and $\overline{p}$ transverse mass spectra are reproduced. 
The emerging trend is that shuffling more flow towards the
fireball edge leads to a greater reduction of $R_{AA}$. The reasons are rather involved:
In order to experience transverse flow at all, the jet production vertex has to be away from the
center of the transverse plane and the hard parton needs to propagate transverse to the local radial
direction $\hat{\bf e}_r$. On the other hand, hard partons produced close to the surface tend to escape before a
significant amount of transverse flow could be generated. In the end, hard partons still in the medium at
relatively late times will be close to the surface and experience additional energy loss if the flow
profile leads to larger flow for the periphery.

\begin{figure*}[!htb]
\epsfig{file=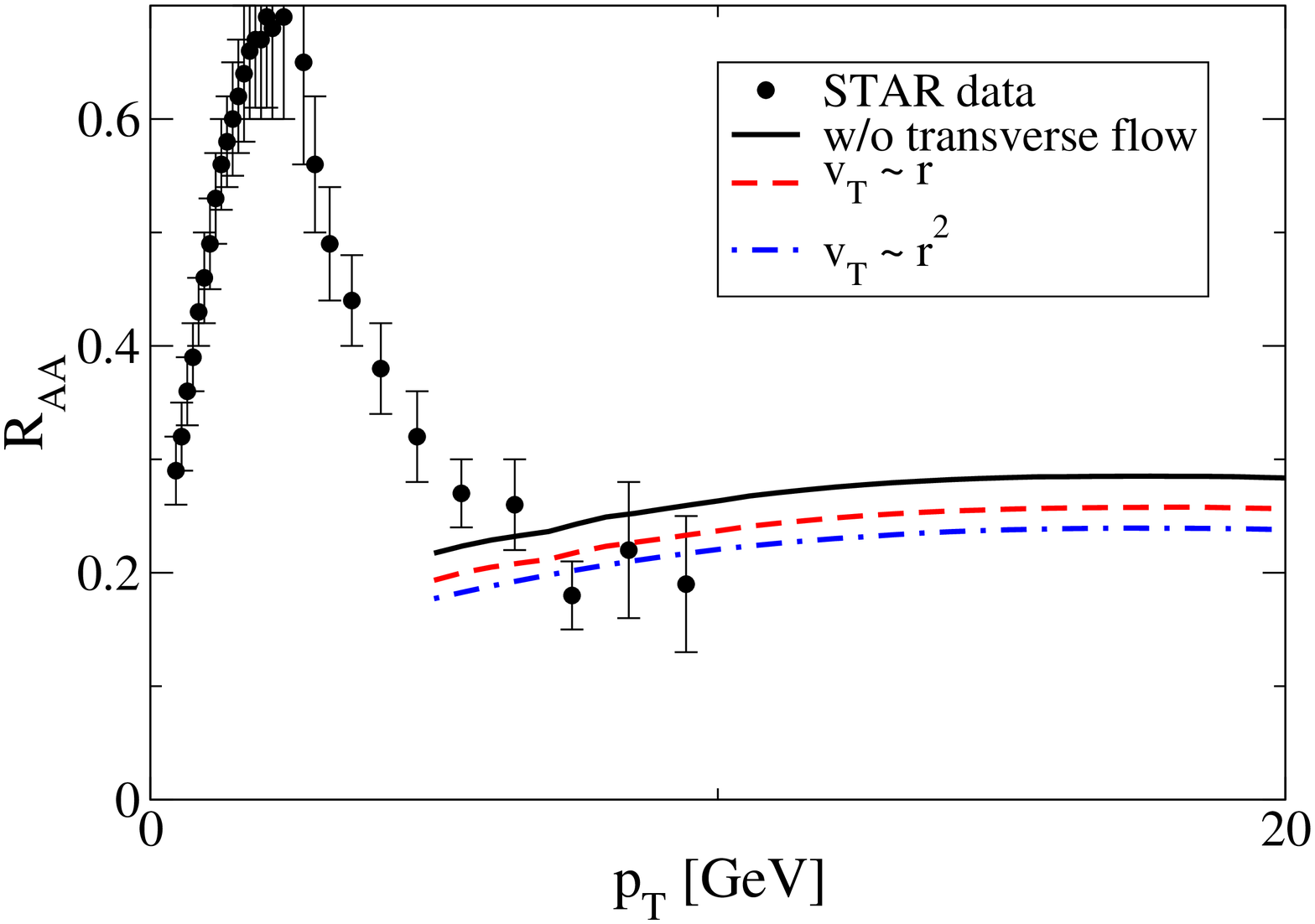, width=8cm}\epsfig{file=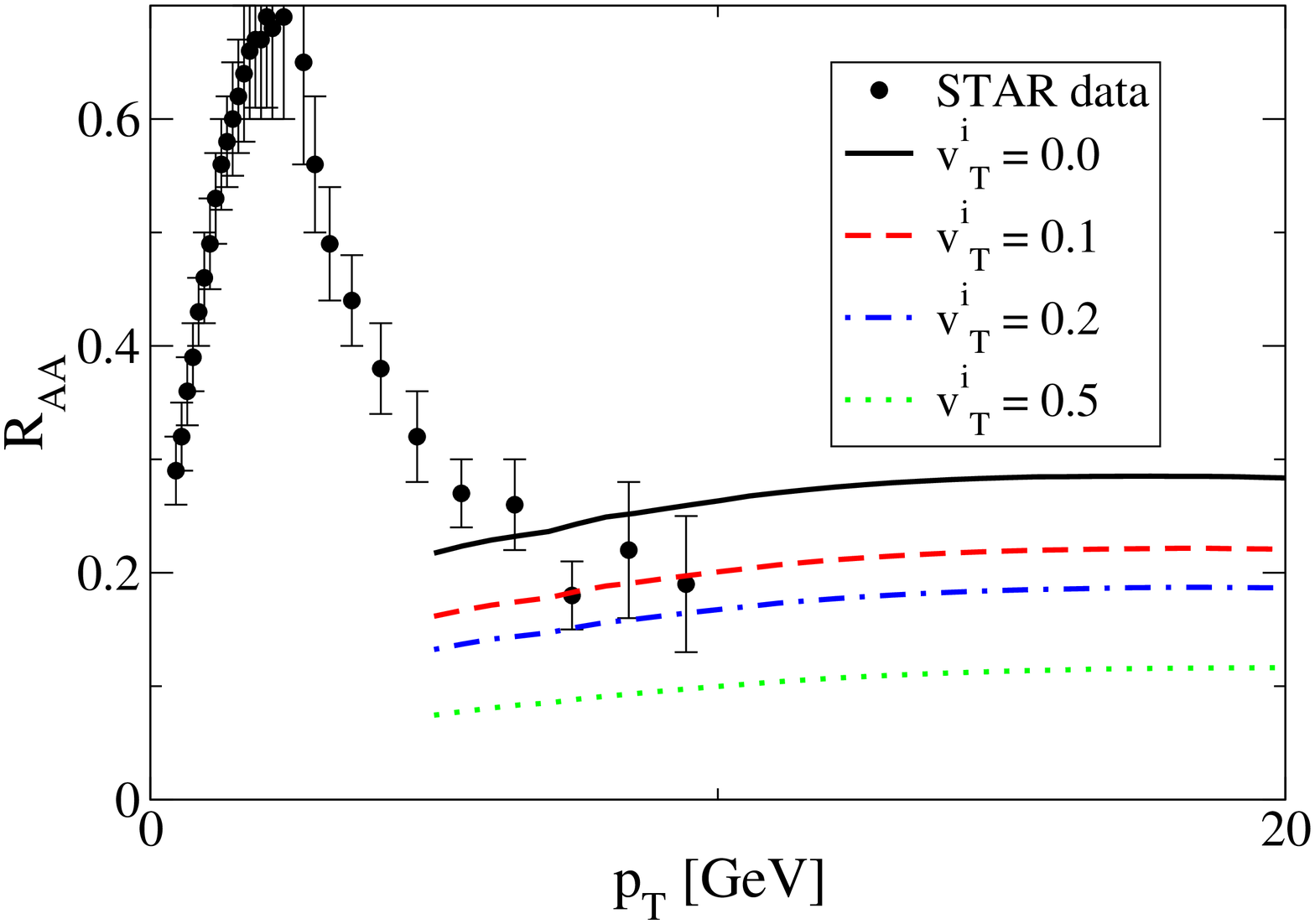, width=8cm}
\caption{\label{F-2}Left panel: Calculated $R_{AA}$ without transverse flow (solid) and with a flow
profile linear (dashed) and quadratic (dash-dotted) in radius as compared to STAR data \cite{STAR_RAA} Right panel: 
Calculated $R_{AA}$ with linear flow profile for different values of pre-equilibrium flow velocity at the fireball edge
$v_T^i$.}
\end{figure*}

The main reason that the influence of transverse flow is much less pronounced than in the findings of 
\cite{Urs2} is that in a realistic evolution scenario the transverse flow field does not build up instantaneously
but rather develops gradually  over time --- as jet energy loss is a phenomenon predominantly sensitive
to earlier times when the medium energy density is large, the spacetime region of the evolution where
transverse flow is small is probed with $R_{AA}$. We demonstrate this in Fig.~\ref{F-2} where
we vary the amount of pre-equilibrium transverse flow $v^i_T$ at the fireball edge. (We take care
that the final transverse flow agrees with the measured $m_T$ spectra by reducing the flow gained during
the thermalized evolution accordingly). The amount of primordial transverse flow is expected to be small,
but our results demonstrate clearly that we recover the comparatively large sensitivity seen in
\cite{Urs2} when we assume that a strong flow field is present {\it ab initio}.

\begin{figure}[!htb]
\epsfig{file=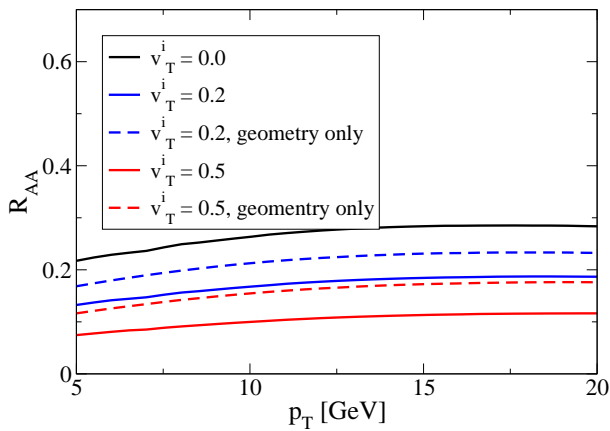, width=8cm}
\caption{\label{F-3}Calculated $R_{AA}$ without primordial transverse flow $v_T^i$ (solid black) and
for $v_T^i$ 0.2 (blue) and 0.5 (red) showing just the geometrical effect of increased radius (dashed) and
 the full effect of geometry and increased energy loss (solid).}
\end{figure}

However, just increasing the amount of primordial transverse expansion is hardly an appropiate comparison as
in addition to the increased role of flow in the energy loss there is a rather trivial geometrical effect:
The fast initial expansion of the fireball radius in the presence of primordial flow 
increases the average pathlength of hard partons in matter.
In order to separate the two effects, we show a calculation in which only the geometrical effect was
taken into account but where we neglected the flow effect on $\hat{q}$.
The results are shown in Fig.~\ref{F-3}.

In essence, for not too large primordial expansion velocities, geometry accounts for about half of the
observed effect. Note that a small primordial flow field cannot be ruled out by current experimental
evidence: One of the observables most sensitive to the initial state is the emission of thermal
photons, but using the framework outlined in \cite{PhotonHBT} we observe that the increased radial flow
tends to compensate the faster cooling by the increased expansion so that it is very difficult to distinguish the
scenarios.

In principle, in a non-boost invariant accelerated longitudinal expansion there can be a component of 
longitudinal flow transverse to the jet axis (this is not so in a Bjorken expansion where the hard parton
is always co-moving with the surrounding matter in longitudinal direction). The magnitude of this 
component depends on the amount of longitudinal acceleration the system has undergone and the distance
of the jet from midrapidity. In the present framework, it turns out that longitudinal flow leads to 
an additional 2\% suppression at $y=0.5$ as compared to a situation in which only transverse flow is
taken into account. This is small compared to other uncertainties, hence we do not discuss the effect 
of longitudinal flow with regard to additional energy loss any further.

\section{The relation between energy density and transport coefficient}

In \cite{Kari1}, an estimate within a Bjorken expansion model was made for the value of the parameter
$c$ relating the transport coefficient and energy density. The analysis found $c > 8 \dots 19$ far from
the perturbative value $c\approx 2$ valid for the ideal QGP. This is in fact consistently larger than
our findings neglecting flow. The reason is that we do not use a Bjorken expansion. This leads to
stronger initial compression of matter (and thus higher densities) as well as slower subsequent
cooling in our model. If we assume a Bjorken expansion and neglect the effect of transverse flow
on energy loss we find good agreement with the data for $c=10$,
quite consistent with the estimate in \cite{Kari1}.

\begin{figure}[!htb]
\epsfig{file=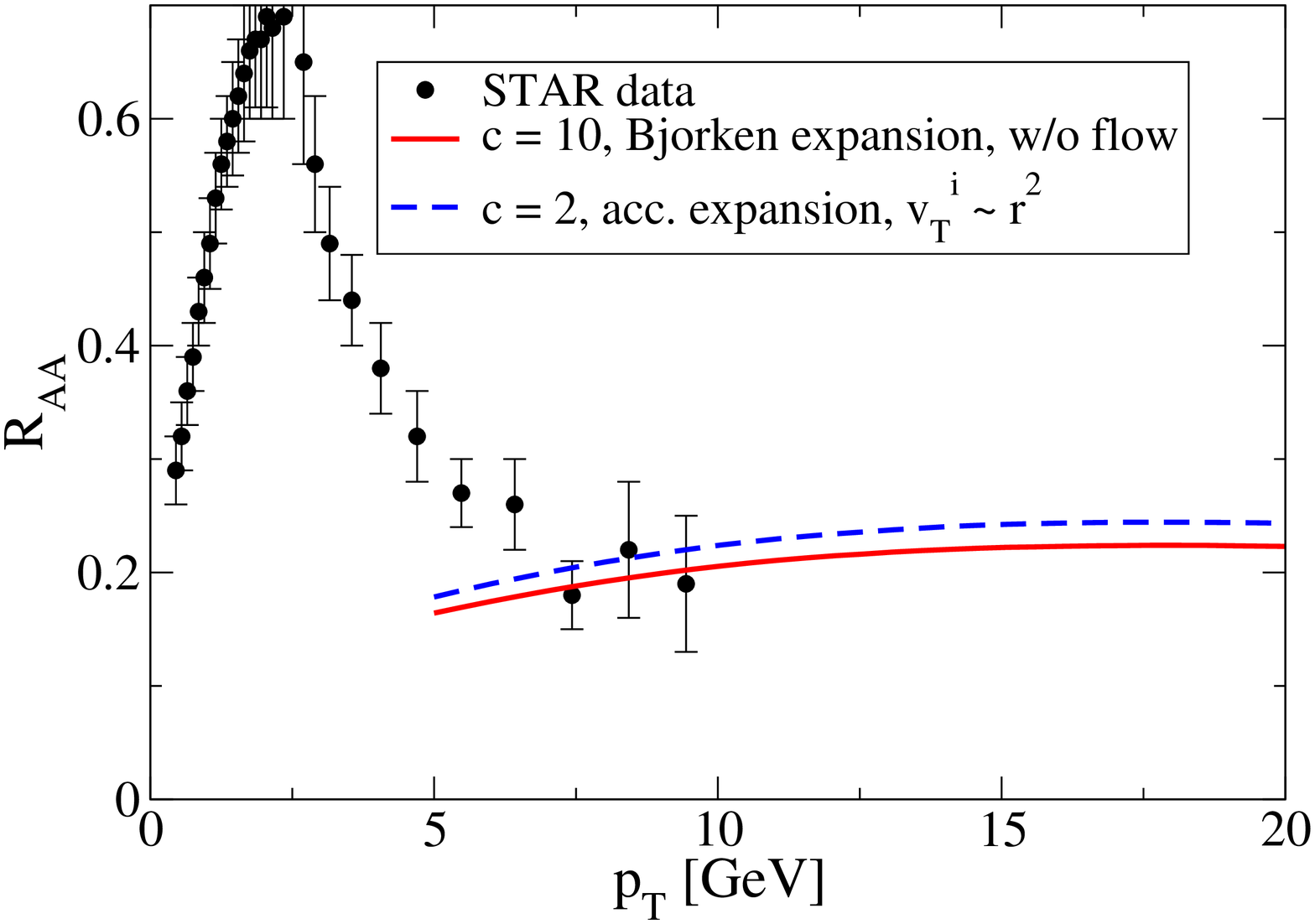, width=8cm}
\caption{\label{F-4}Calculated $R_{AA}$ in a Bjorken expansion and without the effect of transverse flow on
energy loss, requiring $c=10$ (solid red) and in the best fit evolution described in \cite{RenkSpectraHBT}, assuming
a quadratic flow profile, a small primordial flow velocity $v^i_T = 0.1$ and $\alpha_s = 0.45$. 
In this case, only $c=2$ is required for agreement with the
data.}
\end{figure}

However, if we take transverse flow into account in a moderately optimistic scenario, i.e.
assuming a quadratic flow profile, a small value of primordial flow with $v_T^i = 0.1$ and
furthermore increase $\alpha_s$ to 0.45 instead of 0.3 then we find that $c=2$ in fact
gives a fair description of the data. Thus, there may not be a huge quantitative discrepancy between
pQCD predictions and the measured energy loss. This is shown in Fig.~\ref{F-4}.

\section{Summary}

We have investigated the effect of flow transverse to the jet direction on the energy loss of the leading
parton. The net effect arises from an intricate interplay of several different effects, among them
the quadratic pathlength dependence of energy loss which tends to emphasize the role of late times, the expansion and
cooling matter which dilutes the system and emphasizes early times and gradual buildup of the transverse flow field
over time. Jets need to be produced away from the center of the transverse distribution in order to
see a component transverse to the jet direction, but if they are produced too close to the edge
they are likely to escape before flow could build up. As a result, we find great sensitivity to
the presence of a primordial flow field. 

We have also studied the role of a longitudinal flow component orthogonal to the outgoing jet. We observe that away from midrapidity additional
energy loss is induced if the jet is not longitudinally co-moving with the thermalized matter, however around midrapidity this effect turns out to be small compared to the influence of transverse flow.

If we include the additional energy loss induced by transverse flow in a favourable (but still realistic) scenario, 
we find that the ideal QGP relation $\hat{q} \approx 2 \epsilon^{3/4}$ gives a fair description of the
data. Thus, deviations from the ideal QGP energy loss may not be large even in the temperature range
$< 2 T_C$ probed at RHIC energies.

\begin{acknowledgments}
We thank U.~Wiedemann, S.~A.~Bass and B.~M\"{u}ller for helpful discussions, comments and their support during the preparation of this study.

This work was supported by the DOE grant DE-FG02-05ER41367. The authors are supported as Feodor Lynen-Fellows by the Alexander von Humboldt Foundation.
\end{acknowledgments}

\end{document}